\documentclass[10pt,journal,twocolumn,nofonttune]{IEEEtran}
\newcommand{\nc}{\newcommand}
\nc{\bsm}{\boldsymbol}
\nc{\mbb}{\mathbb}
\nc{\mbs}{\mathbbmss}
\nc{\mbf}{\mathbf}
\nc{\mcl}{\mathcal}

\newif\ifhavebib

\usepackage{blindtext}
\usepackage{graphicx}
\usepackage{graphicx}
\usepackage{array}
\usepackage{url}
\usepackage{algorithmic}
\usepackage[cmex10]{amsmath}
\usepackage{ifpdf}
\usepackage{amssymb,amsmath}
\usepackage{empheq}
\usepackage{stfloats}   
\usepackage{textcomp}
\usepackage{cite}
\usepackage{multirow}
\usepackage{diagbox}
\usepackage{subcaption}
\let\oldFootnote\footnote
\newcommand\nextToken\relax

\renewcommand\footnote[1]{%
    \oldFootnote{#1}\futurelet\nextToken\isFootnote}

\newcommand\isFootnote{%
    \ifx\footnote\nextToken\textsuperscript{,}\fi}

\usepackage[ruled,vlined,lined,ruled,boxed]{algorithm2e}
\usepackage[dvipsnames,usenames]{color}
\usepackage[normalem]{ulem}
\usepackage{amsthm}

\usepackage{graphicx}
\usepackage{ifpdf}
\usepackage{cite}
\usepackage{enumerate}
\usepackage{enumitem}

\usepackage{array}
\usepackage{mdwmath}
\usepackage{mdwtab}
\usepackage{eqparbox}
\usepackage{bm}

\usepackage{setspace}
\usepackage{booktabs}
\usepackage[subnum]{cases}
\usepackage{rotating}

\usepackage{listings} 
\definecolor{Red}{rgb}{1,0,0}
\definecolor{Blue}{rgb}{0,0,1}
\definecolor{Olive}{rgb}{0.41,0.55,0.13}
\definecolor{Green}{rgb}{0,1,0}
\definecolor{MGreen}{rgb}{0,0.8,0}
\definecolor{DGreen}{rgb}{0,0.55,0}
\definecolor{Yellow}{rgb}{1,1,0}
\definecolor{Cyan}{rgb}{0,1,1}
\definecolor{Magenta}{rgb}{1,0,1}
\definecolor{Orange}{rgb}{1,.5,0}
\definecolor{Violet}{rgb}{.5,0,.5}
\definecolor{Purple}{rgb}{.75,0,.25}
\definecolor{Brown}{rgb}{.75,.5,.25}
\definecolor{Grey}{rgb}{.5,.5,.5}

\newcommand{\boxhead}[5]{
   \pagestyle{myheadings}
   \thispagestyle{plain}
   \setcounter{page}{1}
   \noindent
   \begin{center}
   \framebox{
      \vbox{\vspace{2mm}
    \hbox to 6.28in { {\bf #1 \hfill} }
       \vspace{6mm}
       \hbox to 6.28in { {\Large \hfill \bf #2  \hfill} }
       \vspace{6mm}
       \hbox to 6.28in { {\it #3 #4 \hfill  #5} }
      \vspace{2mm}}
   }
   \end{center}
   \markboth{#5 -- #2}{#5 -- #2}
   \vspace*{4mm}
}

\usepackage[colorlinks=true, urlcolor=black,  linkcolor=black, citecolor=black]{hyperref}
\theoremstyle{definition}

\theoremstyle{remark}

\theoremstyle{definition}

\DeclarePairedDelimiterX{\infdivx}[2]{(}{)}{%
	#1\;\delimsize\|\;#2%
}

\newcommand{\thetav}{\boldsymbol \theta}

\def\e{\epsilon}

\def\de \mathrm{d}

\newcommand\eg{e.g.,\xspace}
\newcommand\ie{i.e.,\xspace}
\def\textiid{i.i.d.\@\xspace}

\newcommand\iid{\ifmmode\text{ i.i.d. } \else \textiid \fi}

\newcommand{\beqs}{\begin{equation*}}
\newcommand{\eeqs}{\end{equation*}}
\newcommand{\beq}{\begin{equation}}
\newcommand{\eeq}{\end{equation}}
\begin{document}



\havebibtrue
	\author{
		Hang~Liu, Zehong~Lin, Xiaojun~Yuan, and Ying-Jun~Angela~Zhang
\thanks{
H. Liu, Z. Lin, and Y.-J. A. Zhang are with The Chinese University of Hong Kong;	X. Yuan is with the University of Electronic Science and Technology of China. 
}
}

\title{Reconfigurable Intelligent Surface Empowered Over-the-Air Federated Edge Learning
}
\maketitle

\begin{abstract}
Federated edge learning (FEEL) has emerged as a revolutionary paradigm to develop AI services at the edge of 6G wireless networks as it supports collaborative model training at a massive number of mobile devices. However, model communication  over wireless channels, especially in uplink model uploading of FEEL, has been widely recognized as a bottleneck that critically limits the efficiency of FEEL. Although over-the-air computation can alleviate the excessive cost of radio resources in FEEL model uploading, practical implementations of over-the-air FEEL still suffer from several challenges, including strong straggler issues, large communication overheads, and potential privacy leakage. In this article, we study these challenges in over-the-air FEEL and leverage reconfigurable intelligent surface (RIS), a key enabler of future wireless systems, to address these challenges. We study the state-of-the-art solutions on RIS-empowered FEEL and explore the promising research opportunities for adopting RIS to enhance FEEL performance.
	

\end{abstract}

\section{Introduction}

Sixth-generation (6G) wireless communications are envisioned as intelligent information systems that are both driven by and drivers of  artificial intelligence (AI). On one hand, AI will make 6G smart, agile, and able to learn and adapt to the changing network dynamics. On the other hand, the unprecedented capacity and flexibility of 6G will facilitate the deployment of mobile AI services to support diversified computation-intensive mobile applications, such as autonomous driving, auto-robots, and augmented reality. These applications stimulate the development of large-scale machine learning (ML) technologies that can make use of gargantuan amounts of mobile data generated by ever-increasing edge devices.
Moreover, new smart devices exhibit a compelling increase in computation capability, making on-device local training possible for sophisticated AI models. With all that being said, the future wireless networks will be designed to support distributed ML that leverages local resources, including local computation units and data, at the edge level. This calls for revolutionary \emph{communication} and \emph{computation} techniques to meet stringent latency  and accuracy requirements with limited radio resources such as bandwidth and power. 

Edge ML has generated considerable recent research interest; one of the most important approaches is federated learning (FL) \cite{FEDSGD}, which collaboratively trains a unified AI model for different parties without direct data exchange. When FL is deployed at the edge level, known as federated edge learning (FEEL)\cite{GZhu_BroadbandAircomp}, edge devices perform local training using local data and periodically exchange \emph{model information} instead of raw data with a parameter server (PS, usually a base station in a wireless network) to update the global model. 
While FEEL improves the computation efficiency by exploiting the computation capabilities of local devices, frequent model communications between the edge devices and the PS critically limit the performance of FEEL. To see this, note that a FEEL network generally comprises thousands or even millions of edge devices (\eg smartphones and Internet-of-things (IoT) devices). It is shown that model communication, especially in \emph{uplink model uploading}, can be slower than local computation by many orders of magnitude due to limited radio resources.\footnote{This article is focused on the uplink model uploading and aggregation of over-the-air FEEL as these two steps are regarded as the main bottlenecks of FEEL.  Downlink model broadcasting, however, is another interesting topic that deserves dedicated research efforts.
 } Over-the-air computation has been introduced into FEEL to relieve the communication burden of uplink model uploading \cite{GZhu_BroadbandAircomp}. Specifically, over-the-air computation allows massive devices to simultaneously upload local models over the same time-frequency resources  by exploiting the signal superposition property of multiple-access channels. With over-the-air computation, the latency or the required bandwidth in FEEL model uploading is independent of the number of devices and thus can be vastly reduced.

Although over-the-air FEEL is envisioned to be a scalable solution, there still remain three unresolved challenges that limit its communication performance: First, over-the-air FEEL suffers from an inherent \emph{communication-learning tradeoff} induced by device selection. While selecting a subset of edge devices to participate can improve the communication quality, {\color{blue}device selection decreases the number of exploited training data and hence may degrade the learning convergence} \cite{GZhu_BroadbandAircomp}.\footnote{\color{blue} Low-quality data caused by contamination or poisoning attacks jeopardize the learning convergence and should be discarded, \eg by anomaly detection. However, quality-aware data sampling is beyond the scope of this paper.}
Such a conflict puts practical device selection in a dilemma and complicates the design on both communication and learning aspects. Second, the design of over-the-air FEEL critically relies on transmitter-side signal scaling to align the local models coherently at the receiver. This is usually achieved by using the channel state information at the transmitter side (CSIT). However, CSIT acquisition requires frequent channel information feedback from the PS, which  incurs \emph{large communication overheads} and causes a large time delay to FEEL. Third, transmitting local model information reveals sensitive information, which can be used by potential eavesdroppers for privacy attacks.
Enhancing the privacy level, \eg by adding artificial noises,  leads to learning accuracy degradation in FEEL because the added noise makes the estimation of uploaded models at the PS less accurate.

From the above discussions, we see that all the three remaining challenges in canonical over-the-air FEEL come down to the limitation of the volatile wireless propagation environment. This limitation can be potentially eliminated by the forward-looking vision of ``smart radio environment" in future wireless systems. As a key enabler of the smart radio environment, reconfigurable intelligent surface (RIS) perceives wireless propagation channels as programmable entities that can be dynamically configured \cite{yuan2020reconfigurableintelligentsurface}. Specifically, a RIS is a two-dimensional metasurface consisting of a large number of tiny, low-cost, and passive reflecting elements that can induce adjustable phase shifts to the incident signals \cite{wu2020intelligent}. The RIS-enabled wireless networks are thus able to enhances
the channel conditions of devices by tuning the RIS phase shifts.

{RIS-enhanced communications have already attracted widespread interest in the community. {The challenges and developments for RIS-assisted wireless communications can be found in, e.g., \cite{yuan2020reconfigurableintelligentsurface,wu2020intelligent}.}
	While much research has investigated RIS and over-the-air FEEL separately in recent years,
	limited work has been done towards an integrated design on RIS-assisted over-the-air FEEL.  In particular, Ref. \cite{FL_RIS1} made a preliminary investigation on the use of RIS to mitigate the over-the-air model aggregation error. {\color{blue} Ref. \cite{liu2020reconfigurable} reported that RIS can efficiently enhance the model aggregation accuracy and balance the communication-learning tradeoff in over-the-air FEEL. Ref. \cite{liu2020CSITFREE} utilized RIS to reduce the downlink feedback overhead for over-the-air FEEL. These works have demonstrated the importance of RIS in addressing the first two challenges of over-the-air FEEL, namely the communication-learning tradeoff and large communication overheads. Moreover, Ref. \cite{9069945} has shown that the privacy of FEEL model aggregation is determined by the channel conditions of edge devices. RIS is envisioned to address the privacy leakage issue of over-the-air FEEL thanks to its capability of configuring the wireless channels. However, the above literature considers different setups and focuses on resolving only one specific challenge of over-the-air FEEL. To better investigate the effect of RIS on over-the-air FEEL, we need a unified RIS-assisted model communication framework.  
}

}    
{\color{blue}Motivated by this, we aim to spur research attention towards the use of RIS in resolving communication bottlenecks of over-the-air FEEL in this article.} Different from the existing magazine article \cite{FL_RIS1} that mainly aims to introduce RIS into FEEL systems, we focus on adopting RIS for addressing the three major communication challenges in over-the-air FEEL, namely the communication-learning tradeoff, the large overhead in CSIT feedback, and the privacy leakage issue. Specifically, we investigate a unified RIS-empowered FEEL framework, review state-of-the-art efforts in addressing these challenges under this framework, and discuss implementation issues of the current design as well as promising research opportunities. 

The remainder of this article is organized as follows. {\color{blue}In Section II, we review over-the-air  FEEL and point out the three major challenges therein. In Section III, we introduce the RIS technique and propose a unified RIS-empowered FEEL framework to address these challenges.} In Sections IV-VI, we discuss the state-of-the-art solutions to each research challenge under the proposed framework. {\color{blue}In Section VII, we discuss other research opportunities.}
Finally, we conclude this article in Section VIII. 

\section{\color{blue}Over-the-Air FEEL}\label{sec2}
 \begin{figure}[!t]
 		\centering
	\includegraphics[width=3.2in]{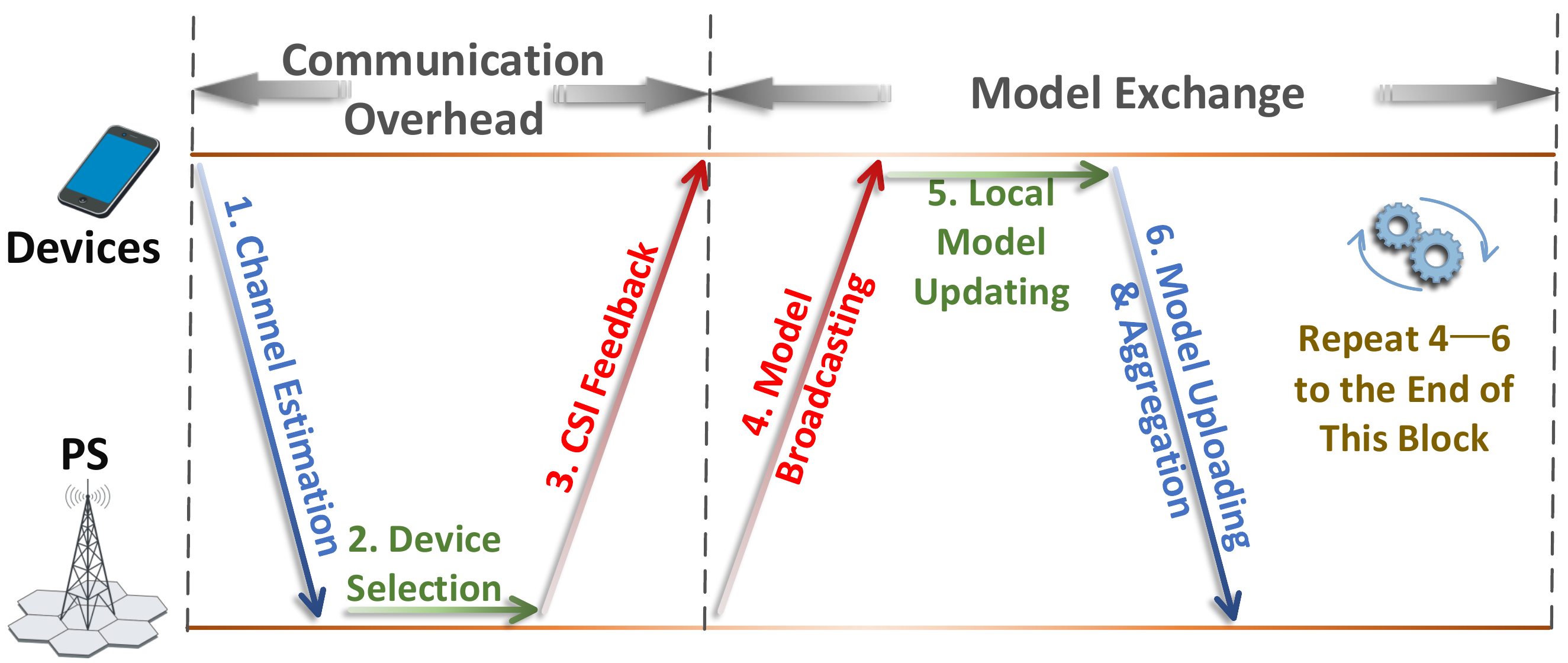}
 	
	\caption{\small The FEEL workflow.}
	\label{figsystem}
\end{figure}

A general FEEL system comprises a PS and a number of edge devices  in a wireless network, with each device possessing a local training dataset and establishing a channel link with the PS. 
The edge devices aim to learn a uniform AI model from the local data. That is, the goal is to find an optimal parameter set of a given AI model by minimizing an additive {empirical loss function with respect to all local training samples}; see \cite{FEDSGD}.  

In FEEL, each device {trains a local model that minimizes the local loss function using its on-device data samples.} Furthermore, the PS iteratively exchanges model information with the devices until convergence. 
Moreover, the acquisition of the latest channel state information (CSI) is critical in FEEL system design. In this regard, in each coherence block we dedicate a number of symbols, a.k.a. communication overhead, for channel training before model transmission. As shown in Fig. \ref{figsystem}, FEEL in a coherence block involves the following steps:
\begin{enumerate}
	\item Edge devices send training pilots to the PS to estimate CSI.
	\item The PS selects a subset of devices to participate into the learning process. 
	\item The PS feedbacks the CSI to the device through downlink control channels.
	\item The PS broadcasts the global model to the selected devices.
	\item Each selected device updates the local model by using, \eg stochastic gradient descent.
	\item The selected devices upload the local gradients or model changes {defined as the element-wise difference between the initial model and the trained local model} to the PS. {The PS
	aggregates a weighted sum of the local model changes and updates the global model.}
\end{enumerate}
Steps 4--6 are repeatedly executed until the end of the coherence block, and the next block begins with Step 1. 
\subsection{\color{blue}Over-the-Air Model Aggregation}
 \begin{figure}[!t]
	\centering
	\includegraphics[width=3in]{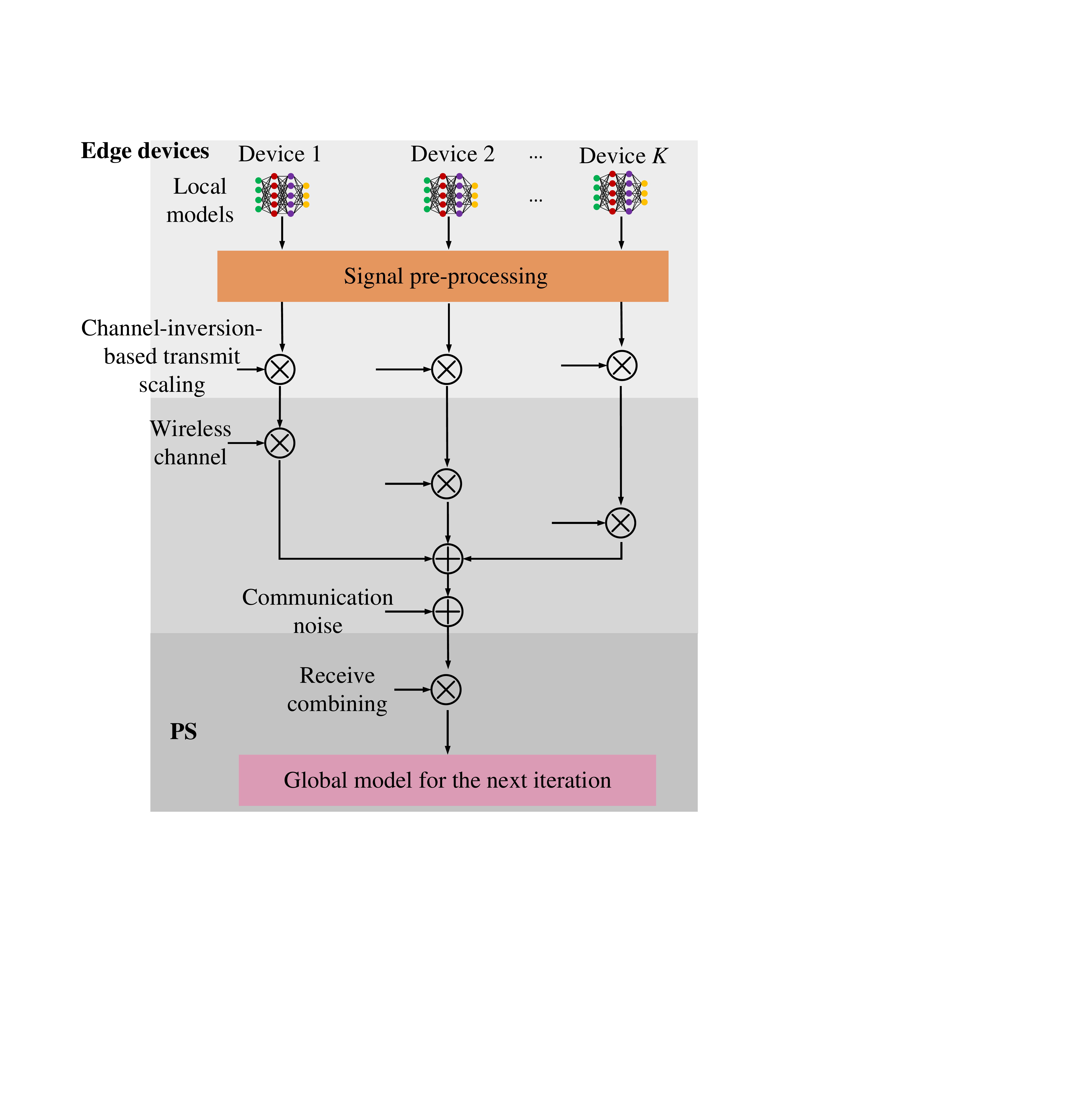}
	\caption{\small \color{blue} The over-the-air model aggregation framework.}
		\label{figsystem2}
\end{figure}
A federated learning network generally comprises numerous edge devices. Consequently, concurrently uploading such a massive number of local models through multiple-access wireless channels costs a large amount of radio resources and incurs an excessively long delay. As a result, model uploading and aggregation in Step 6 have been widely recognized as the main bottleneck of FEEL \cite{FEDSGD}. To tackle the challenge in Step 6,
over-the-air computation has been introduced to support 
 a large number of simultaneous model uploading \cite{GZhu_BroadbandAircomp}. {\color{blue} The over-the-air computation process is illustrated in Fig. \ref{figsystem2}. In each learning iteration, edge devices transmit local model changes over the same physical channel by following the channel inversion principle. 
Specifically, each device sets the transmit scaling coefficient as the ratio of the desired model aggregation weight over its instantaneous channel coefficient. By doing this, the channel fading coefficients are canceled at the PS, and the received signal at the PS is a noisy version of the desired linear combination of local models due to the superposition property of the wireless channel.
 	 Then, the PS estimates the weighted sum of the local models as the updated global model for the next training iteration.
As the required bandwidth or latency does not scale with the number of edge devices, the over-the-air computation is deemed to be a scalable solution to model aggregation.}

\subsection{Major Challenges in Over-the-Air FEEL}\label{seccha}
Although over-the-air FEEL alleviates high bandwidth burden in model aggregation, there are still three key challenges yet to be adequately addressed.  We elaborate on these challenges here, and motivate our design on RIS-empowered FEEL in the next subsection.

First, the performance of over-the-air model aggregation is limited by the communication-learning tradeoff. Specifically, wireless channel conditions vary significantly across mobile devices. Consequently, a  straggler issue exists in the sense that the model aggregation performance is dominated by the devices with weak channel qualities, a.k.a. the communication stragglers.  This is because the devices with better channel qualities have to lower their transmit power so that signals from all devices are coherently aligned at the PS. In order to limit the model aggregation error, the aforementioned stragglers have to be excluded from model training through device selection in Step 2 of Fig. \ref{figsystem}. However, selecting a subset of devices to participate may slow down the learning convergence \cite{GZhu_BroadbandAircomp}. Furthermore, in light of heterogeneous data distributions among edge devices, excluding the stragglers exacerbates the convergence rate loss as it causes a bias to the global model aggregation. 
\begin{figure}[!t]
	\begin{minipage}[t]{0.49\linewidth}
	\centering
	\includegraphics[width=1.5 in]{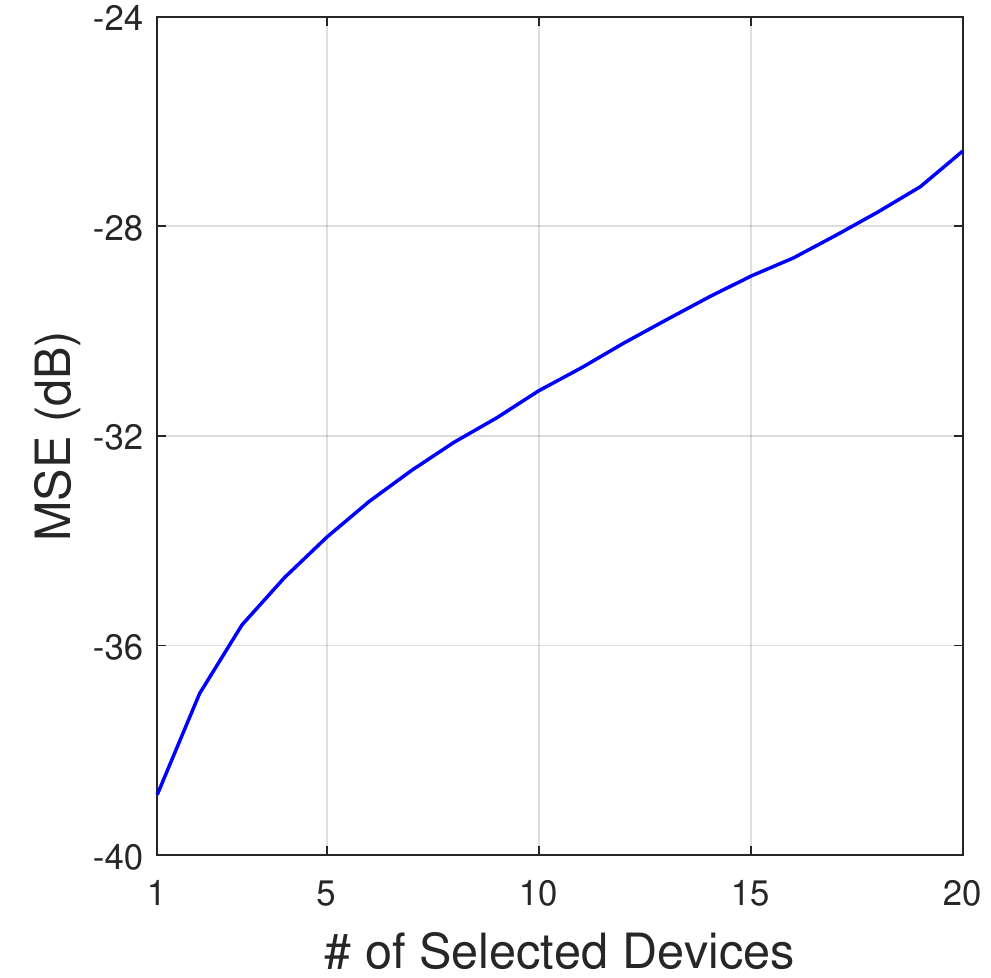}
	\subcaption{}
	\label{figa}
\end{minipage}
\begin{minipage}[t]{0.49\linewidth}
	\centering
	\includegraphics[width=1.5 in]{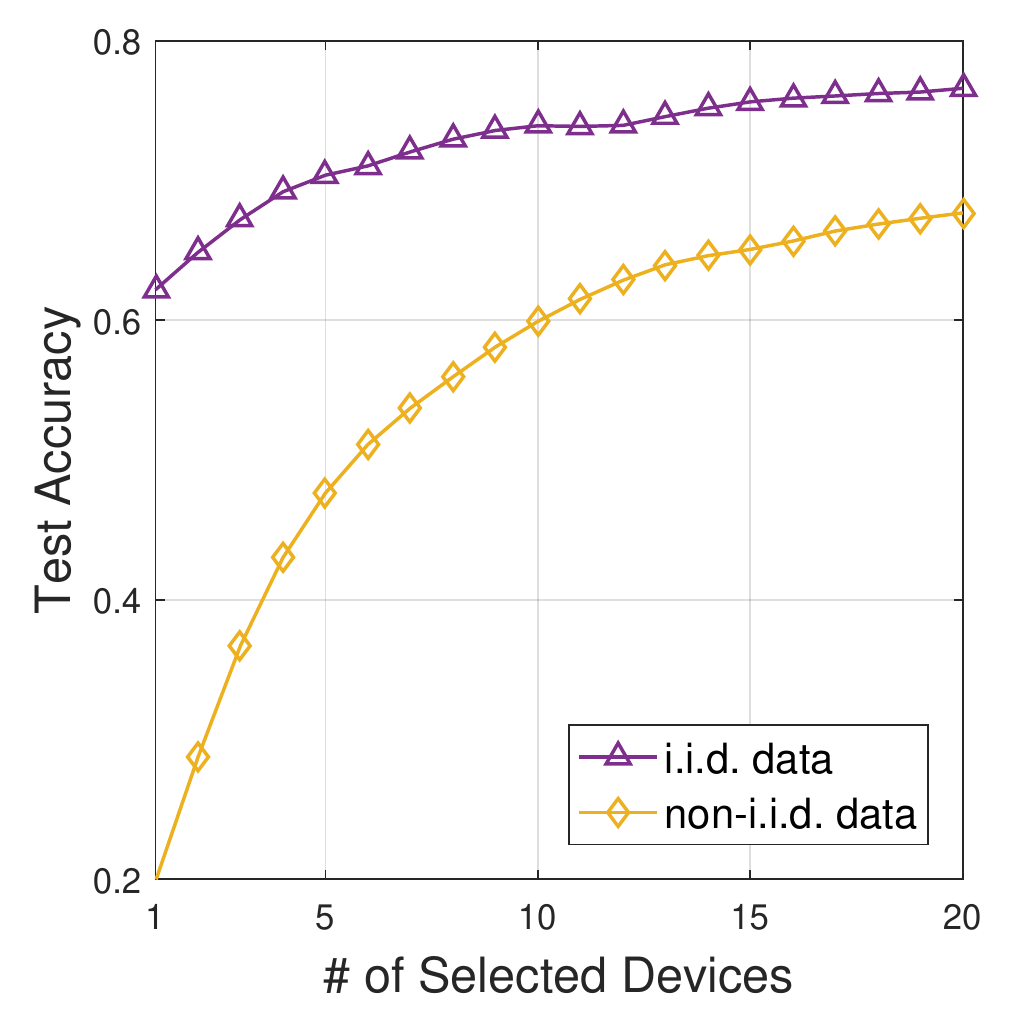}
	\subcaption{}
	\label{figb}
\end{minipage}%
\caption{\small Tradeoff between communication and learning in device selection with $20$ devices. In subfigure (a), devices with i.i.d. Gaussian channels transmit standard Gaussian signals. The devices are selected in the \emph{descending} order of their channel gains. In  subfigure (b),  the learning model in \cite[Sect. V]{liu2020reconfigurable} is simulated {\color{blue}over the Fashion-MNIST dataset} with \emph{error-free} channels, where each device has $100$ samples. 
}
	\label{fig2}
\end{figure}

The existence of the stragglers leads to a fundamental tradeoff between minimizing the communication error and maximizing the number of participants in device selection. This is referred to as the communication-learning tradeoff in over-the-air FEEL\footnote{{The communication-learning tradeoff in traditional FEEL usually refers to the tradeoff between communication \emph{delay} and learning performance \cite{FEDSGD}. In contrast, the communication-learning tradeoff in over-the-air FEEL refers to the tradeoff between model communication \emph{error} and the convergence rate, where the two aspects have a coupled impact on the learning performance.}}, and is
illustrated in Fig. \ref{fig2}. On one hand, Fig. 3(a) shows that the model aggregation  mean-square-error (MSE) significantly increases when more devices participate in training. On the other hand, if the communication error is intentionally neglected by assuming an error-free channel, the learning test accuracy  (in the range of $[0,1]$) improves with more devices selected. Compared with the i.i.d. data distribution case, accuracy degradation from device discarding becomes more severe when data distributions are non-i.i.d.; see Fig. 3(b). 
In order to optimize the learning accuracy, the conflict in minimizing the communication error and maximizing the test accuracy with respect to device selection should be well balanced in a unified optimization framework. However, capturing the impact of device selection in learning performance loss is difficult as there is no direct relationship between the final learning accuracy and the device selection decision for a general learning model. Moreover, the mixed impact of device selection (Step 2 in Fig. \ref{figsystem}) and communication error in model aggregation (Step 6) makes the problem more difficult to tackle.
We summarize this  challenge in Challenge C1.
\begin{description}
	\item[C1] {\color{blue}The joint optimization of the device selection and the communication scheme is challenging due to the communication-learning tradeoff in over-the-air FEEL.}
	 
\end{description}

Second, to facilitate over-the-air model aggregation, the PS needs to feedback CSI to the selected devices for transmitter-side scaling; see Step 3 in Fig. \ref{figsystem}.\footnote{Alternatively, the PS can compute and transmit the scaling factors to the devices. This incurs the same signaling overhead as sending CSI to the devices.}
In order to compute the transmit scaling coefficients, the most up-to-date {CSI} is needed at each device. 
The frequent update of the transmit scaling coefficients incurs high feedback costs in Step 3 of Fig. \ref{figsystem}. Furthermore, due to the limited bandwidth of downlink control channels, CSI is usually quantized and compressed before being transmitted, which brings additional error and leads to imperfect signal alignment in over-the-air model aggregation. This challenge is summarized as follows.
\begin{description}
	\item[C2] To facilitate coherent model aggregation, the PS needs to frequently feedback CSI to the devices. This incurs high downlink signaling overhead and extra signal alignment error.
\end{description}
 \begin{figure*}[!t]
	\centering
	\includegraphics[width=5.2in]{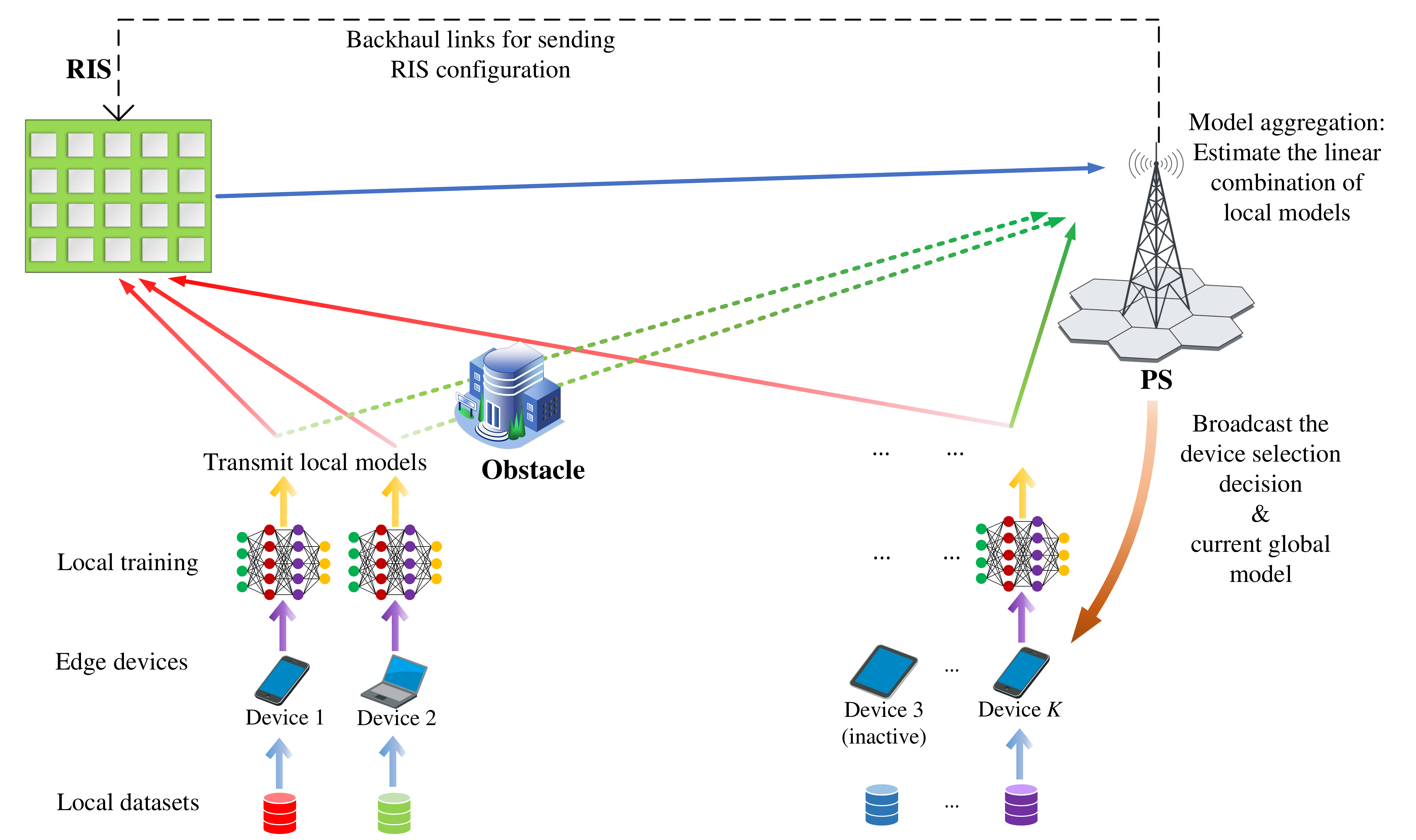}
	\caption{\small An illustration of the RIS-empowered over-the-air FEEL framework in the $t$-th learning iteration. The direct channel links of Devices $2$ and $3$ are blocked by an obstacle, and the RIS is employed to enhance their channel conditions by creating reflecting links.}
	\label{RIS}
\end{figure*}

The last challenge lies in the potential privacy leakage in FEEL. 
{\color{blue}Since over-the-air FEEL needs to share local model information with the PS in the model aggregation process, communicating local models reveals sensitive information and is prone to privacy attacks from an untrusted server. For example, batch-averaged gradients transmitted in learning iterations can be reverse engineered to recover the corresponding training images. Consequently, the private information about local data can be leaked in model aggregation.}
Therefore, privacy-preserving mechanisms, such as differential privacy (DP), secure multiparty computation, and encryption, are needed to protect the privacy for FEEL. However, these mechanisms often achieve data privacy at the cost of reduced learning performance or additional communication resources. For example, a widely adopted approach in DP is to add independent artificial noises to the local model updates before transmission. This inevitably brings errors to model aggregation and degrades the learning performance. {\color{blue} The DP level in over-the-air FEEL is determined by the minimum channel gain between the active users and the receive antennas, resulting in a complicated learning-privacy trade-off and a more challenging optimization problem for over-the-air FEEL with privacy constraints.}
We summarize the challenge in privacy leakage as follows. 
\begin{description}
	\item[C3] Model information exchanged during the training process leaks sensitive information, raising privacy concerns. Privacy-preserving design, on the other hand, often provides privacy at the cost of reduced model performance. {The privacy leakage in over-the-air FEEL is determined by the minimum channel gain, making the system design challenging.}
\end{description}

\section{Overview on RIS-Empowered Over-the-Air FEEL}
To tackle Challenges C1--C3, communication mechanisms in conventional communication systems should be revisited and radio resources should be specifically optimized with respect to FEEL objectives, such as training loss or training time, other than conventional communication metrics. This not only complicates the transceiver design but also increases the radio resource requirements. For example, to address the communication-learning tradeoff, an additional device selection procedure, \ie Step 2 of Fig. \ref{figsystem}, is required before model transmissions, which increases the computational complexity of system optimization. 
Finally, in order to enhance the privacy preservation in FEEL by DP, more transmit power is required to introduce artificial noises.

We envision that the above detriments brought in addressing Challenges C1--C3 can be effectively alleviated by an emerging wireless communication technique called \emph{RIS}. Specifically,  
a RIS
can shape the incident signals by inducing independent phase shifts and can be configured in real time under the control of a central controller. 
{\color{blue}With RIS, the wireless channel can be configured to achieve diverse system requirements at a low cost by optimizing the RIS phase shifts. To this end, it is shown in \cite{yuan2020reconfigurableintelligentsurface,wu2020intelligent} that the RIS phase shifts should be jointly optimized with the transceiver design with respect to the metric of the specific communication application. Moreover, optimizing the RIS phase shifts needs CSI on both the direct and reflecting channels. This requires new  low-overhead  channel estimation and CSI feedback protocols, \eg by exploiting channel statistics or the line-of-sight (LoS) availability; see \cite{yuan2020reconfigurableintelligentsurface} and the references therein.}
Particularly, RIS can be adopted to address Challenges C1--C3 in FEEL model aggregation without incurring high signal processing complexity nor high power/energy consumption, as detailed in the remaining of this article.
We here propose a RIS-empowered FEEL framework as shown in Fig. \ref{RIS}. The RIS is adopted in assisting the model uploading from the edge devices to the PS. Specifically, by introducing the RIS to FEEL, new device-RIS-PS channel links are created to enhance the existing direct device-PS channel links to improve the channel quality. Furthermore, the newly added device-RIS-PS channel coefficients can be  tuned by adjusting the phase shifts induced by the RIS elements. 
Consequently, the overall channel of each device can be regarded as a function of the RIS phase shift vector $\thetav$ and be configured by optimizing $\thetav$. 

{Each training round of the proposed RIS-empowered over-the-air FEEL has the following steps. First, the PS selects the active devices in the next round and broadcasts the current model to them. {\color{blue}The RIS configurations are optimized at the PS and sent to the RIS controller to implement RIS phase shifts\footnote{\color{blue}The RIS phase shifts can be optimized at the PS and sent to the RIS controller via the separate backhaul links \cite{wu2020intelligent}.}}. Meanwhile, the transmit scaling factors at the devices are computed by using the superposed channel coefficients. Then, the devices transmit their model changes to the PS via over-the-air computation so that they are appropriately added at the PS.}
Due to the communication-learning tradeoff, the transceiver design, the RIS phase shifts, and the device selection should be jointly designed under a unified metric that characterizes both the communication and learning impacts on FEEL. {Fig. \ref{RIS} illustrates a typical scenario of RIS-empowered over-the-air FEEL. The PS intends to exploit the model information at several edge devices to maximize the learning convergence rate. However, such devices suffer from large model aggregation errors due to the bad channel conditions of the direct channel links. With the help of the RIS, we can create extra reflection links to enhance the channel conditions of those communication stragglers, and hence the dilemma in the communication-learning tradeoff can be broken.} Moreover, we note that the proposed RIS-empowered FEEL framework unifies the existing designs on RIS-enabled FEEL \cite{wang2020federated,ywliu,liu2020reconfigurable,liu2020CSITFREE}. We shall review the state-of-the-art solutions on RIS-empowered FEEL in addressing Challenges C1--C3 in the subsequent sections.

\section{RIS for Communication-Learning Co-Design in  Over-the-Air FEEL}\label{sec3}
As discussed in Section \ref{seccha}, the channel heterogeneity across edge devices leads to the communication-learning tradeoff that fundamentally limits the over-the-air FEEL performance. 
RIS 
can be used to enhance the reliability of wireless channels, as most prior work on RIS-assisted communications does, so that more devices can participate in the training process without causing noticeable degradation of the model aggregation performance. Furthermore,  the RIS phase shifts can be co-designed with the communication variables (\ie the transceiver design) and the learning scheme (\eg device selection) so that RIS can be better leveraged to empower FEEL. {Due to the characteristic of over-the-air model aggregation, the learning performance depends on the channel conditions of the communication stragglers, rendering an objective different from conventional communication systems. Therefore, new  designs on RIS phase shifts are needed for over-the-air FEEL.} In this section, we review the state-of-the-art solutions on RIS-empowered communication-learning co-design and discuss their limitations.

Refs. \cite{wang2020federated,ywliu} optimized the RIS design and the device selection by maximizing the number of selected devices under a predefined communication error constraint.  
Both works reported that RIS-assisted communication-learning co-design leads to noticeable convergence improvement compared with the one without RIS.
%
\begin{figure}[!t]
	\begin{minipage}[t]{0.49\linewidth}
		\centering
		\includegraphics[width=1.6 in]{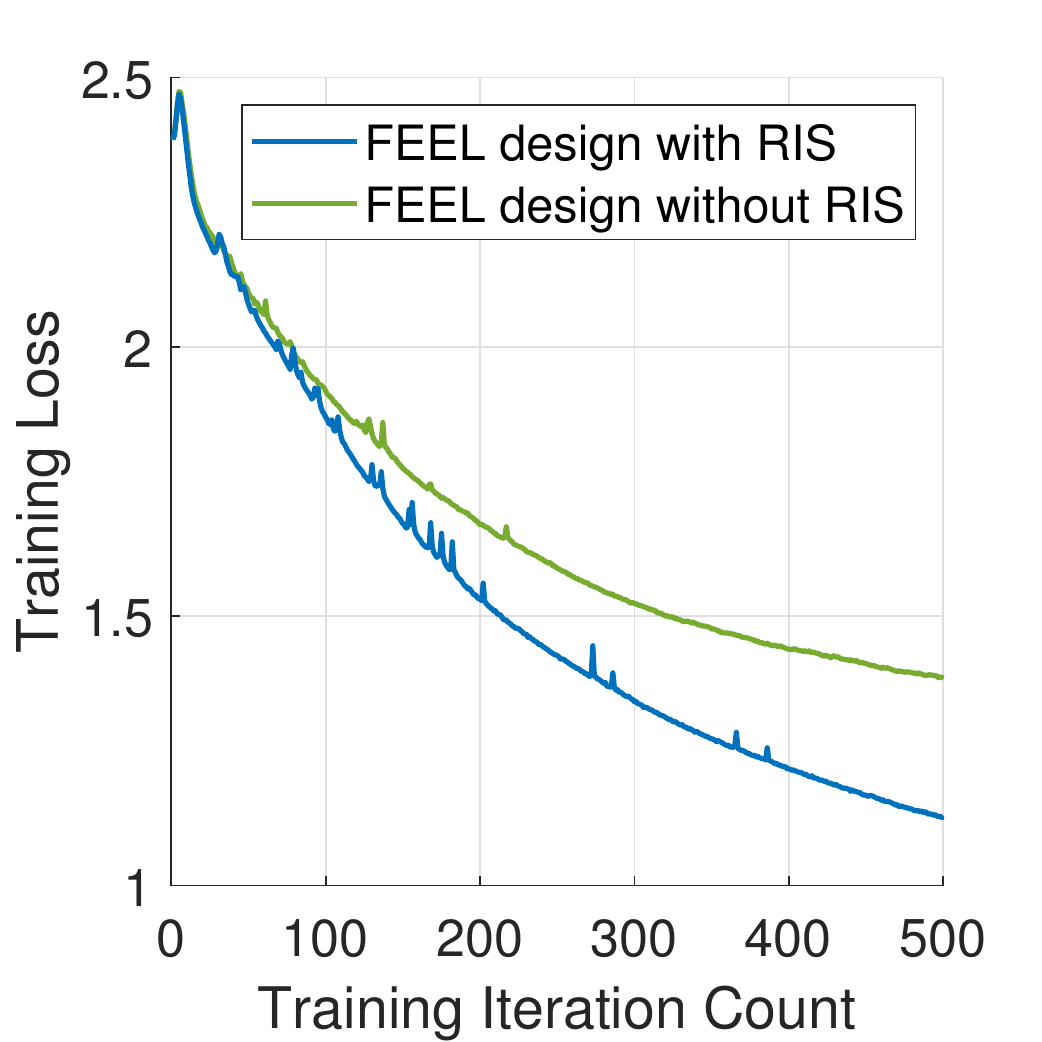}
		\subcaption{}
		\label{fig3a}
	\end{minipage}
	\begin{minipage}[t]{0.49\linewidth}
		\centering
		\includegraphics[width=1.6 in]{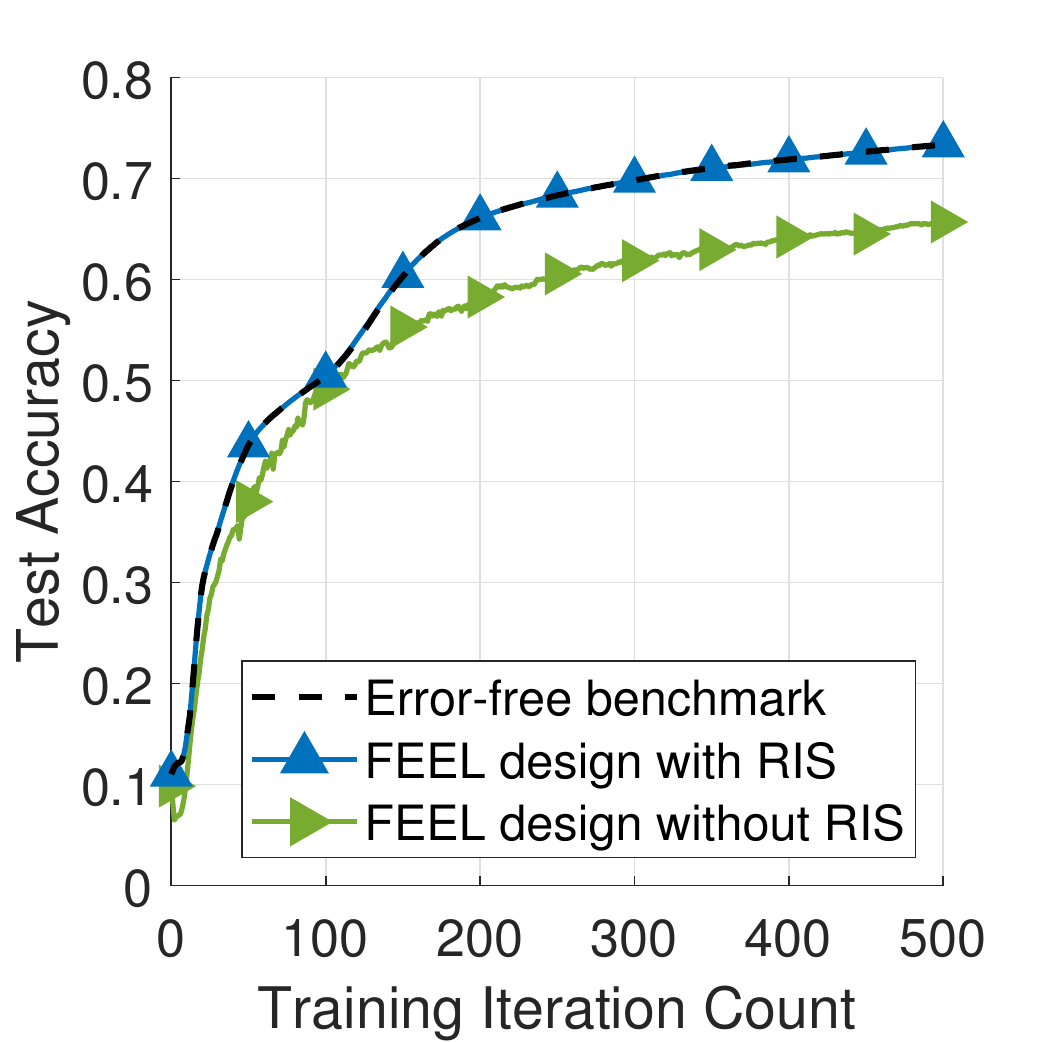}
		\subcaption{}
		\label{fig3b}
	\end{minipage}%
	\caption{\small {Training loss (left) and test accuracy (right) of the communication-learning co-design  \cite{liu2020reconfigurable}.} The \emph{error-free} benchmark characterizes the best possible learning accuracy with error-free channels. The simulation parameters are the same  as those in \cite[Fig. 4]{liu2020reconfigurable}.}
	\label{fig3}
\end{figure}
Moreover, Ref. \cite{liu2020reconfigurable} designed RIS-empowered FEEL by directly analyzing the coupled impacts of both aggregation error and device selection loss. 
It shows that the training loss
is upper bounded by a weighted sum of a device selection loss and a communication loss: 
While the device selection loss decreases when more devices are selected, the loss resulted from communication error increases when devices with weak channels are selected. Furthermore, Ref. \cite{liu2020reconfigurable} jointly optimize the RIS phase shifts, the receiver beamforming, and the device selection by directly minimizing the overall learning performance loss. 
Fig. \ref{fig3} plots the learning accuracy of the communication-learning co-design   approach in \cite{liu2020reconfigurable} with or without RIS. 
{\color{blue}The RIS phase shifts efficiently enhance the channel conditions and well balance the communication-learning tradeoff. As a result, the model aggregation error of the RIS-assisted design is sufficiently small, which does not affect the gradient descent directions in the training process.}
This result demonstrates {\color{blue} the importance of  RIS in enhancing the communication-learning co-design in FEEL}.

{\color{blue}The above approaches suffer from high computational complexity in the high-dimensional optimization of the large RIS phase-shift vector.  Possible low-complexity substitution of high-dimensional optimization is a codebook-based RIS design. For example, we can design a RIS phase shift codebook with each codeword beamforms the incident signal to different directions and search for the best  RIS configuration codeword. 
Moreover, recent studies have shown that the RIS design complexity  can be significantly reduced by exploiting the channel statistic information in conventional wireless networks \cite{9198125}. We envision that the channel statistics can also be applied to facilitate the low-complexity communication-learning co-design in over-the-air FEEL. 
Furthermore, we highlight two possible extensions of the existing RIS-empowered communication-learning co-design. First, the current work in \cite{wang2020federated,liu2020reconfigurable,ywliu} assumes fixed device selection in each coherence block. Note that time-varying device scheduling (\ie selecting different devices in different iterations)  is generally more favorable than the fixed one as the former can exploit diversified training data on different devices. However, time-varying device scheduling significantly complicates the design of RIS-empowered FEEL. 
Second, as shown in Fig. 3(b), the device selection loss is more severe when data are non-identically distributed across devices. In this case, the RIS plays a more profound role in improving the channel qualities of devices with important data so that they do not need to be discarded in device selection. To achieve this goal, system optimization should also consider the impact of the data heterogeneity.}

	 \begin{figure}[!t]
	\centering
	\includegraphics[width=2.5in]{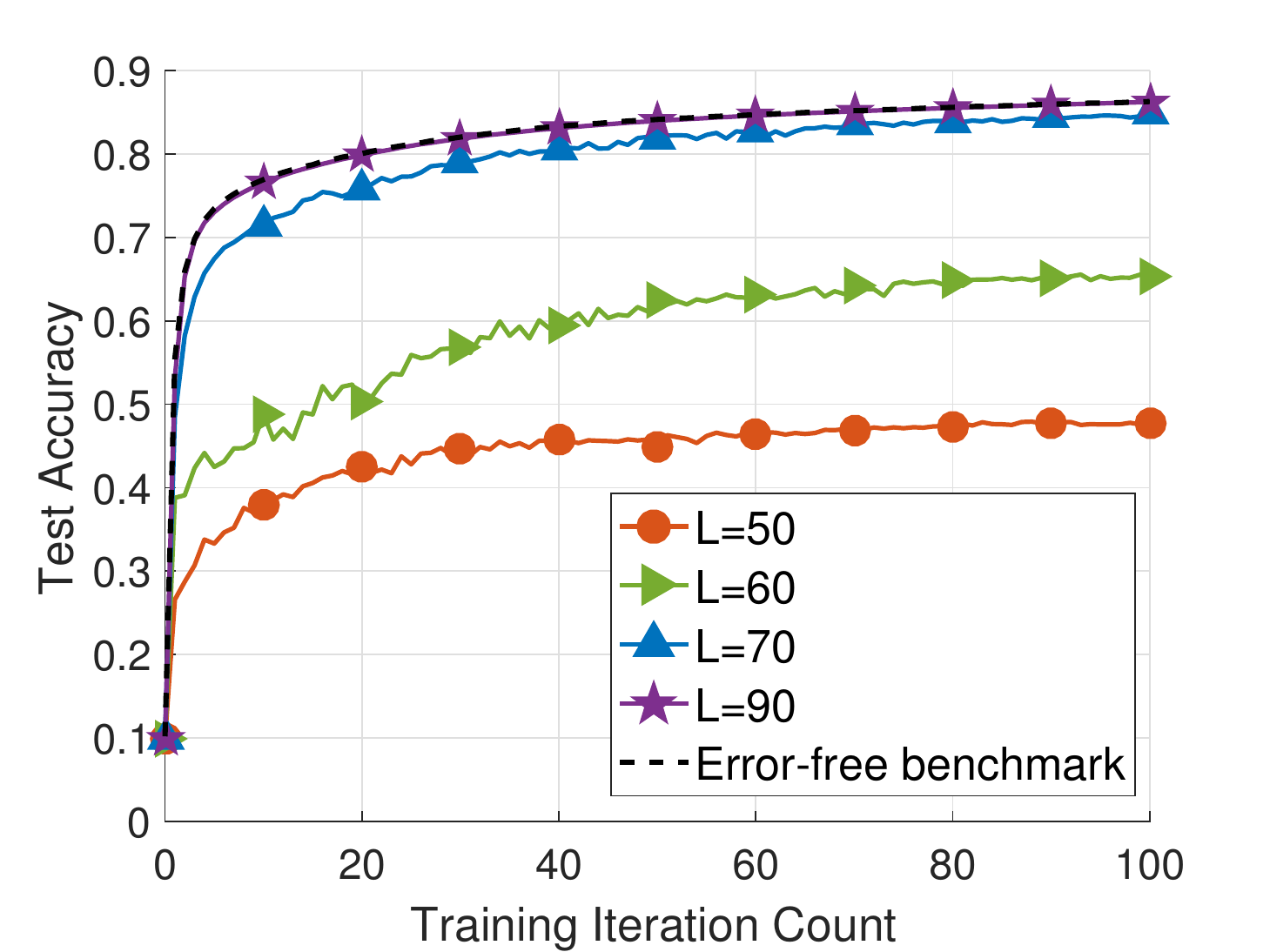}
	\caption{\small Test accuracy of the RIS-empowered CSIT-free FEEL algorithm in \cite{liu2020CSITFREE} with $40$ edge devices, where $L$ denotes the number of RIS elements. \emph{Error-free} benchmark is also included, which assumes no communication error. The method in \cite{liu2020CSITFREE} achieves a near-optimal test accuracy when $L\geq 90$.}
	\label{fig4}
\end{figure}
\section{RIS for Low-Overhead Over-the-Air FEEL}\label{sec4}
The approaches in Section  \ref{sec3} all assume perfect CSIT for transmit scaling. 
However, acquiring CSIT incurs high communication overhead in Step 3 of Fig. \ref{figsystem}. Furthermore, the inevitable error in CSI quantization/communication degrades the signal alignment accuracy in model aggregation.
Massive multiple-input multiple-output (MIMO) has been utilized to avoid the CSI feedback by achieving CSIT-free transmit scaling \cite{amiri2020blind}, but it increases power consumption and deployment costs. 
{\color{blue} As RIS can significantly improve the energy efficiency of wireless communications \cite{yuan2020reconfigurableintelligentsurface}, it is necessary to leverage RIS for achieving CSIT-free FEEL.} 
A recent study in \cite{liu2020CSITFREE} utilized RIS to design CSIT-free FEEL systems. Specifically, it is assumed that no CSIT is available at the devices, and the PS is only equipped with one receive antenna. Consequently, the authors in \cite{liu2020CSITFREE} 
configured the RIS phase shifts $\thetav$ so that the resultant channel coefficient of each device is approximately proportional  to the corresponding aggregation weight. In this way, 
the local model updates are coherently aligned through adjusting channel coefficients via the RIS. 
Fig.  \ref{fig4} plots the performance of the CSIT-free FEEL approach in \cite{liu2020CSITFREE} with various numbers of RIS elements $L$. 
{\color{blue}With a sufficiently large RIS, the RIS-assisted design achieves a sufficiently small model aggregation error, whose effect on the gradient descent direction is negligible. Therefore, it achieves a similar accuracy to the error-free benchmark, demonstrating the power of a passive RIS in attaining low-overhead FEEL.} 

We note that the CSIT-free FEEL design mentioned above focuses on the communication aspect only without considering device selection. As discussed in Section \ref{sec3}, communication-learning co-design can further unleash the potential of RIS in empowering FEEL, which remains an open problem in the field of CSIT-free FEEL. We envision low-overhead FEEL with joint device selection and communication design as a promising future research direction. However, the  communication-learning co-design for CSIT-free FEEL becomes more challenging than the CSIT-based one since device selection critically affects the capability of signal alignment by the RIS phase shifts.  Specifically, aligning dissimilar channel coefficients to the predetermined weight  is more difficult and thus usually leads to a large aggregation error. Therefore, {device selection should not only balance the communication-learning tradeoff but also activate devices with  channel coefficients as similar as possible  by, e.g., channel-aware user grouping. Furthermore, the current RIS-empowered design in \cite{liu2020CSITFREE} relies only on the RIS for signal alignment with a single receive antenna, which is prone to CSI error. Combining the MIMO technique and the RIS phase shift optimization shall significantly improve the robustness of the RIS-empowered CSIT-free FEEL design.}

\section{RIS for Private  Over-the-Air FEEL }\label{sec5}

In this section, we study Challenge C3, namely privacy leakage in FEEL.
A widely adopted privacy preserving approach, known as  DP, is for each device to hide its information, \eg by adding artificial noise. 
The analyses in \cite{9069945,9252950} reveal the fundamental tradeoff between learning performance and privacy level: More artificial noise leads to a higher privacy level but compromises the learning performance of FEEL. 
{As discussed in Section \ref{seccha}, DP in over-the-air FEEL has a more complicated expression compared with traditional FL, which makes the privacy enhancement of over-the-air FEEL more challenging. 
	{\color{blue}Considering that the PS is equipped with multiple antennas, the achievable DP at the PS  is given by the minimum DP level at the receive antennas, which is determined by the minimum channel gain among all the transmit-receive antenna pairs. 
	Since RIS can efficiently manipulate wireless channels, it is necessary to explore RIS to enhance the DP in model aggregation.
	Specifically, the role of RIS in DP-aware over-the-air FEEL is to manipulate the wireless channel condition so that  the achievable DP is sufficiently high and, at the same time, the accurate model aggregation is achieved.}
Analogously to the communication-learning co-design in Section \ref{sec3}, the communication (\ie transceivers and RIS phase shifts), learning (\ie device selection), and privacy (\ie the power of artificial noise and the information retrieval mechanism) aspects shall be jointly optimized in a unified optimization framework. A possible formulation is to minimize the expected training loss subject to $\e$-DP privacy preservation, \ie the privacy leakage levels at all the receive antennas are below some given threshold $\e$. To this end, the impacts of the communication-learning-privacy factors on both the training loss and the privacy leakage shall be investigated, and the learning behavior and the DP level shall be modeled with respect to all the considered variables.

\section{\color{blue}Other Opportunities in RIS-Empowered Over-the-Air FEEL}
{\color{blue}
Except for Challenges C1--C3, RIS can also be employed to enhance over-the-air FEEL in the following aspects:
\begin{itemize}
	\item AI-based low-complexity RIS configuration: Optimizing the RIS phase shifts is extremely challenging when the number of RIS elements becomes large. AI technology is efficient in solving such a large-scale optimization problem. Particularly, the deep reinforcement learning (DRL) technique has been adopted for sum-rate maximization in conventional RIS-assisted wireless networks \cite{9110869}. Since the relationship between the RIS design and the learning convergence is difficult to model for general non-convex learning tasks, we envision that data-driven AI techniques such as DRL are promising for low-complexity system optimization in RIS-assisted over-the-air FEEL.

	\item RIS deployment optimization: The performance of RIS-assisted FEEL systems critically depends on the deployment strategy of the RIS, which is overlooked in the related literature. To combat the large path loss of the cascaded RIS reflecting channels, we need to place the RIS in the region where the LoS  links are available for both the device-RIS and the RIS-PS channels. Moreover, since the over-the-air model aggregation error is dominated by the communication straggler with the worst channel condition, the RIS should be placed near such a straggler. In summary, the RIS deployment should simultaneously consider the above factors and be jointly  optimized with other variables such as device selection in the system design.
	\item RIS for passive beamforming and data transfer: Recent studies show that RIS can be integrated with sensors to actively sense the data from the environment. Accordingly, we can use RIS to assist over-the-air FEEL and simultaneously gather training data from the environment. It has been verified that sharing a small amount of additional training data among the edge devices can largely enhance the convergence rate of FL. 
	The passive beamforming and information transfer (PBIT) technique \cite{9117136} can be extended to over-the-air FEEL systems to achieve RIS-assisted model aggregation and data sharing from the RIS to the edge devices. 

\end{itemize}
%
%
}
\section{Conclusions}


This article discussed the fundamental communication challenges in over-the-air FEEL systems: the communication-learning tradeoff, the large communication overhead, and the privacy leakage issue.  We introduced a RIS-empowered FEEL framework to address these challenges and discussed the promising future directions  under this framework.
 We hope this article will spur widespread interest in the integration of FEEL and RIS.


%

\ifCLASSOPTIONcaptionsoff
\newpage
\fi
\ifhavebib
{
	\bibliographystyle{IEEEtran}
	\bibliography{ref}
}
\else{
}
\fi

\textbf{Hang Liu} [S'19, M'21] (hangliu@nus.edu.sg) is a research fellow with the National University of Singapore.

\textbf{Zehong Lin} [S'17] (lz018@ie.cuhk.edu.hk) 
is pursuing the Ph.D. degree in information engineering at The Chinese University of Hong Kong.

\textbf{Xiaojun Yuan} [S'04, M'09, SM'15] (xjyuan@uestc.edu.cn)
is a professor with the University of Electronic Science and Technology of China.

\textbf{Ying-Jun Angela Zhang} [S'00, M'05, SM'10, F'20] (yjzhang@ie.cuhk.edu.hk) 
is a professor with the Department of Information Engineering, The Chinese University of Hong Kong.

\end{document}